\documentclass{PoS}

\usepackage{url}

\newcommand{\matx}{\left |\mathcal{M} \right|^2}
\newcommand{\gev}{{\ensuremath\rm GeV}}

\title{Charged Higgs production with a top in MC@NLO}

\ShortTitle{Charged Higgs production with a top in MC@NLO}

\author{\speaker{Tilman Plehn}\\
        Institut f\"ur Theoretische Physik,\\
        Universit\"at Heidelberg, Germany\\
        E-mail: \email{plehn@uni-heidelberg.de}}
\author{Carole Weydert\\
        Laboratoire de Physique Subatomique et de Cosmologie, UJF,\\
        CNRS/IN2P3, \\
        Grenoble, France\\
        E-mail: \email{weydert@lpsc.in2p3.fr}}

\abstract{The production in association with a top quark is the most
  promising search channel for charged Higgs bosons at the LHC. We
  review its theoretical description including next-to-leading order
  corrections and the combination with a parton shower. The latter
  allows us to for the first time answer questions about the
  kinematics of all jets in the process. We then describe the
  consistent subtraction of intermediate states and present new
  results about the bottom mass uncertainty impacting the parton
  densisities.}

\FullConference{Prospects for Charged Higgs Discovery at Colliders \\
                 27-30 September 2010\\
                 Uppsala University, Sweden}

\begin{document}

\section{Four and five flavors}

As has been known for a long time~\cite{bottom_partons,eduard}
processes relying on gluon splitting into an initial state $b\bar{b}$
pair can be described by generating a universal bottom content in the
proton via the DGLAP evolution and including it as a parton density.
In our case this means that to leading order we compute the process
$p_5p_5 \to t H^-$ with five quark flavors inside the
proton~\cite{charged_th,mine1,mine2,mcnlo_th}. Following
Figure~\ref{fig:feyn} we can as well limit the quark content of the
proton to four light flavors and only generate initial state bottom
quarks via gluon splitting, so our leading order process is $p_4p_4
\to \bar{b} t H^-$~\cite{charged_bth}. The index of the proton
indicates that this implies different proton pictures, {\sl i.e.}
different parton densities and $\alpha_s$ values.

\begin{figure}[b]
\begin{center}
\includegraphics[width=0.5\hsize]{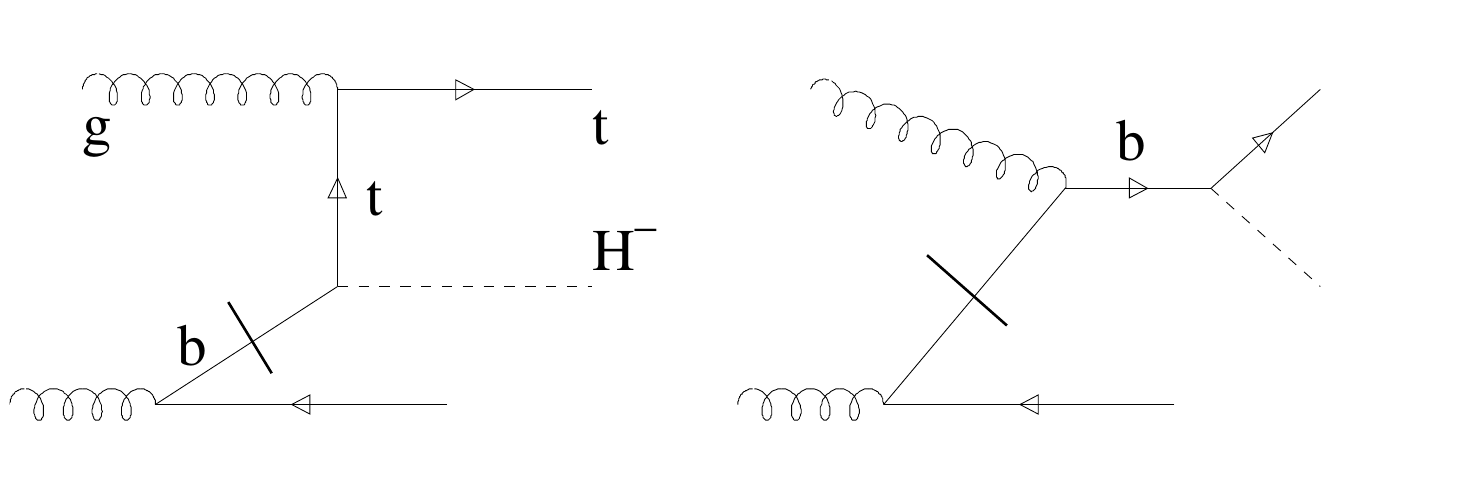}
\end{center}
\vspace*{-10mm}
\caption{Feynman diagrams for associated top-Higgs production at
  the LHC.}
\label{fig:feyn}
\end{figure}

To all orders in perturbation theory the two approaches are
equivalent. The difference between them is the ordering in
perturbative QCD. The DGLAP equation resums collinear logarithms which
appear at every fixed order in $\alpha_s$. Since they diverge for
massless quarks there is no question that for those the logarithms
have to be resummed, to obtain finite and stable rate predictions. For
bottoms the logarithms have the form $\log (M/m_b$) with the process
dependent hard scale $M \sim m_t + m_H$. They are regularized and
reside in the perturbative regime $m_b \gg \Lambda_{\rm QCD}$.

Because of these logarithms, comparisons at leading order have to be
taken with a grain of salt. The renormalization and factorization
scale dependences in both processes are large and the predicted rates
correspondingly uncertain. In the five-flavor scheme the factorization
scale dependence of the total rate is approximately logarithmic and
has a one-to-one correspondence to the collinear
logarithm~\cite{bottom_partons}. Setting the factorization scale to
$\mu_F \equiv M$ implicitly assumes the existence of large logarithms
all the way to $\log (M/m_b)$. A careful study of the kinematics of
gluon splitting generating the logarithms shows that bottom densities
combined with an incoming gluon or bottom never generate as large
logarithms, due to the steep drop of the gluon parton densities
towards large $x$~\cite{eduard,mine1,scott}.  Only a consistent
(lower) choice of the bottom factorization scale ensures a stable
perturbative behavior of the five-flavor production rate. For $t H^-$
production these effects are not huge, while they completely change
the balance of the different production modes for example in the case
of charged Higgs pair production~\cite{charged_pairs}. Moreover, given
that we have fully understood NLO results for the
five-flavor~\cite{mine1,mine2,mcnlo_th} and
four-flavor~\cite{charged_bth} descriptions of $t H^-$ production we
do not expect a huge discrepancy between the different rate
predictions.

Expanding the argument to distributions there appear slight
differences between the two calculations. The distributions of the
heavy top and Higgs are computed with the same perturbative accuracy
as the respective total rates.  Luckily, we know for example from the
production of supersymmetric particles~\cite{prospino} that the
distributions of heavy particles produced in pairs are remarkably
stable with respect to higher orders in QCD, so again we do not expect
large discrepancies.

\section{MC\@@NLO and jet radiation}

The main difference between the NLO four-flavor and five-flavor
calculations appears in the bottom jet distributions.  On the one
hand, the motivation for using bottom partons is that in most events
we will not see the forward and comparably soft bottom jet, and
integrating over its phase space produces the logarithmic rate
enhancement.  On the other hand, modern analysis methods are able to
make use of such features which are only visible in a fraction of the
events, so our simulation has to describe them correctly.

\begin{figure}[t]
\begin{center}
\includegraphics[width=0.6\hsize]{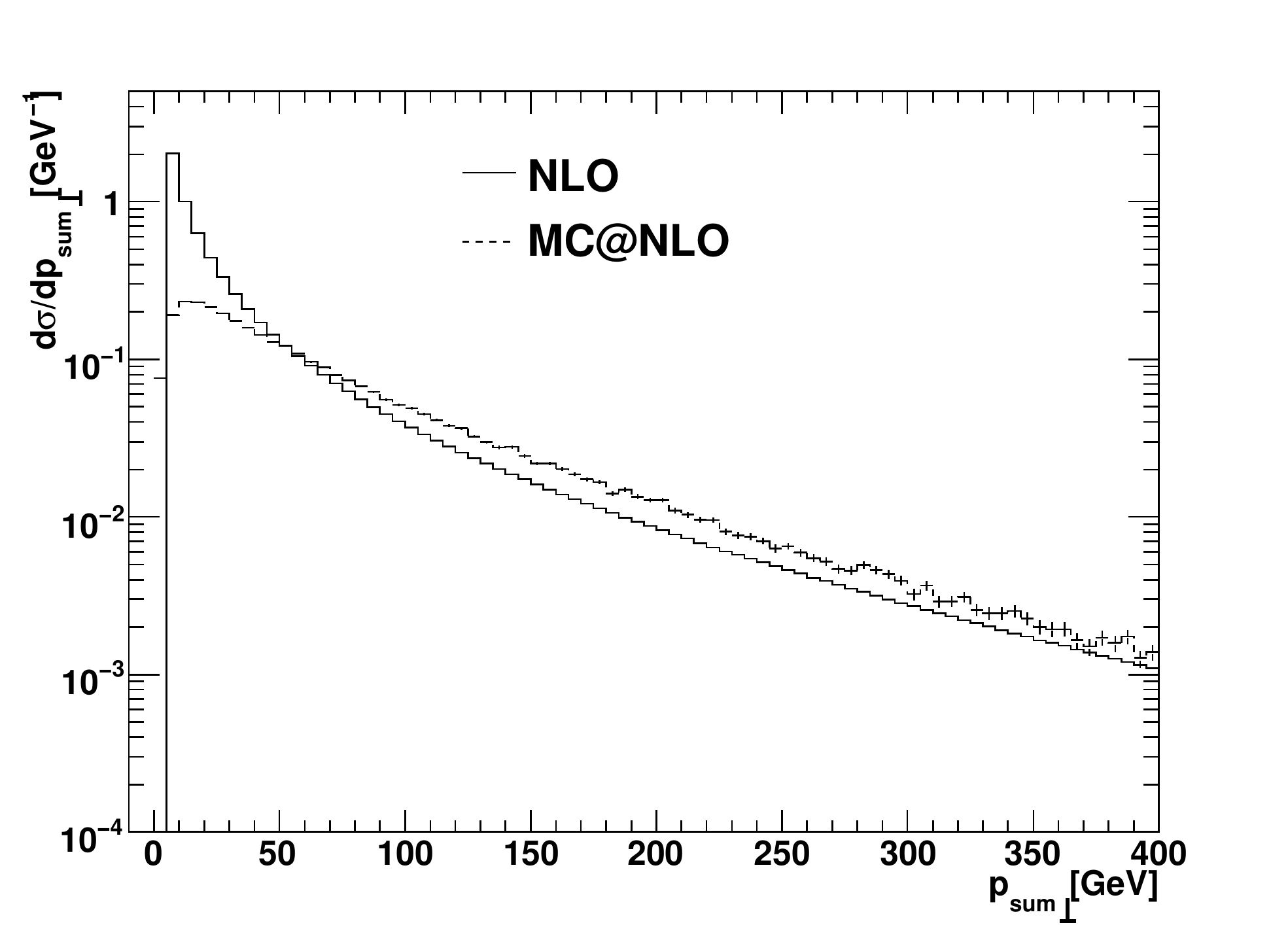}
\end{center}
\vspace*{-8mm}
\caption{Transverse momentum distribution of the top-Higgs pair,
  simulated at fixed order (NLO, five flavors) and including the
  parton shower~\cite{mcnlo_th}. The assumed charged Higgs mass is 300~GeV.}
\label{fig:ptsum}
\end{figure}

To leading order, in the five-flavor scheme the heavy pair recoils
against a strictly longitudinal bottom. Only at NLO this recoiling jet
acquires a finite $p_{T,b}$~\cite{mine2}.  In the four-flavor scheme
the recoiling bottom is realistically described already at leading
order.  At first glance this looks like a case for a four-flavor
calculation, at least when looking at distributions. However, from the
moderate size of the collinear logarithm we know that $p_{T,b} \ll M$,
{\sl i.e.}  in the bulk of its phase space the bottom jet is correctly
described by a parton shower and not by a hard matrix element. What we
should really do is combine either the four-flavor or the five-flavor
description with a parton shower.  This is precisely what MC\@@NLO
does. The impact of the parton shower becomes obvious in the $p_T$
distribution of the top-Higgs system shown in
Figure~\ref{fig:ptsum}. At fixed order it diverges while the resummed
collinear logarithms in the parton shower correctly describe the
low-$p_T$ regime.

Combining the parton shower with the matrix element to NLO ensures
that we correctly describe all jet radiation while the total rate and
the hard distributions are at NLO accuracy.  This way we do not need
an external normalization, as it used to be the case for PYTHIA and
HERWIG. With a built-in NLO rate prediction we can answer questions
of the kind~\cite{mcnlo_th}:

\noindent $\bullet$ What are the transverse momentum or rapidity
spectra of the radiated $b$ jet?

\noindent $\bullet$ Are these $b$ spectra the same as for
light-flavor initial state radiation?

\noindent $\bullet$ Is the $b$ jet from the top decay always harder
than the $b$ jet from initial state radiation?

\noindent $\bullet$ How likely is it that we observe or tag this
radiated $b$ jet?

\noindent $\bullet$ How often do we see light-flavor or $b$
jets given one $b$ jet from the top decay has been observed?

\begin{table}[t]
\begin{small} \begin{center}
\begin{tabular}{lc||c|c|c|c|c||c|c|c|c|c|}
&&\multicolumn{5}{c||}{$m_H = 300$~GeV}
 &\multicolumn{5}{c|}{$m_H = 800$~GeV}\\
\hline
& $p_T^{\rm min} \diagdown \eta^{\rm max}$
& 2.5 & 2.0 & 1.5 & 1.0 & 0.5
& 2.5 & 2.0 & 1.5 & 1.0 & 0.5\\
\hline
($b$ jet, $t_\ell$)
& 25~\gev &17.8  &14.3  &10.0  &5.7  &2.3 
     & 21.4 &17.1  &12.1  &7.1  &3.0 \\
& 45~\gev & 12.9 &10.6  &7.6  &4.5  &1.8 
     &16.6  &13.7  &9.9  &6.0  &2.5 \\
& 65~\gev &9.4  &8.0  &5.9  &3.5  &1.6 
     &12.3  &10.5  &7.9  &4.9  &2.0 \\
& 85~\gev & 7.2 &6.4  &4.8  &3.0  &1.4 
     &9.7  &8.5  &6.5  &4.1  &1.7 \\
\hline
(light jet, $t_\ell$)
& 25~\gev &45.9  &40.0  &32.7  &23.9  &13.0 
     & 54.8 &48.8  &41.0  &31.0  &17.9 \\
& 45~\gev & 32.4 & 27.8 &22.3  &16.1  &9.0 
     & 41.7 &36.7  &30.5  &23.0  &13.7 \\
& 65~\gev &22.3  &18.8  &14.7  &10.4  &5.8 
     &30.9  &27.0  &22.2  & 16.5 &10.2 \\
& 85~\gev &16.2  &13.4  &10.3  &7.3  &4.2 
     & 23.6 &20.5  &16.6  &12.1  &7.4 \\
\hline
(light jet, $t_h$)
& 25~\gev &94.9 &91.0 &84.3  &72.2  &48.4 
     &95.8  &92.5  &86.3  &75.0  &52.0 \\
& 45~\gev &83.2  &79.2  &72.3  &61.0  &39.9 
     &87.1  &83.3  &76.8  &65.7  &45.2 \\
& 65~\gev &60.9  &57.3  &51.7  &43.2  &28.8 
     &70.5  &66.9  &61.3  &51.9  &35.9 \\
& 85~\gev &44.4  &41.5  &37.1  &31.1  &21.3 
     &56.2  &53.3  &48.6  &41.0  &28.7 
\end{tabular}
\end{center} \end{small}
\vspace*{-5mm}
\caption{Probability (\%) to see a $b$ or light jet in a given
  detector region, after one $b$ jet, most likely from the
  top decay, is observed. The top decays leptonically or hadronically
  while the charged Higgs is stable~\cite{mcnlo_th}.}
\label{tab:bfrac}
\end{table}

As an example, we answer the last question in
Table~\ref{tab:bfrac}~\cite{mcnlo_th}. The starting point, motivated
by the usual analyses, is that one bottom jet from the top decay
should always be observed. The additional bottom jet arises from
initial state gluon splitting. In the five-flavor scheme to leading
order the probability of seeing such a bottom jet within the fiducial
volume of the detector is zero. Including the NLO corrections with a
reliable $p_{T,b}$ spectrum over the entire range indicates that in
roughly 18\% of the events we can tag the radiated bottom jet given
$p_{T,b} > 25$~GeV and $|\eta_b| < 2.5$ for $m_H = 300$~GeV. Combined
with typical tagging efficiencies this translates into roughly 10\% of
the inclusive $tH^-$ events with a $b$ tag. Because of the collinear
enhancement the radiated bottom jet becomes harder for larger Higgs
masses, but the generic picture of roughly 10\% of all events
including a tagged forward bottom does not change.

In addition, we can ask if the appearance of a forward jet can be
viewed as a sign for initial-state gluon splitting into two
bottoms. Table~\ref{tab:bfrac} indicates that for a leptonic top
decay, {\sl i.e.} no additional hard jets in the event, it is two to
three times as likely to observe a light-flavor jet from initial state
or final state radiation than a bottom jet. When we in addition allow
for jets from a hadronic top decay almost all events will see an
additional jet with $p_{T,j} > 25$~GeV and $|\eta_j| < 2.5$. This
percentile only drops once $p_{T,j}$ reaches the Jacobian peak of the
top decay.

Again, only the combination of a hard NLO matrix element with the
parton shower allows us to systematically compare such kinematic
features of bottom jet radiation, light-flavor jet radiation, and top
decay jets, as shown in Table.~\ref{tab:bfrac}~\cite{mcnlo_th}.

\section{Bottom mass and bottom partons}

Bottom parton densities are based on the splitting of an off-shell
gluon into a pair of massive bottoms. While for light-flavor quarks
the splitting threshold is of the order of $\Lambda_{\rm QCD}$ and
hence not numerically relevant, for bottoms it is in the range of
perturbative QCD. This makes it a relevant input parameter in the
computation of bottom parton densities, which has to be
tested~\cite{pdf_mb}. To first approximation, a shift in the bottom
mass changes the logarithmic parton densities by

\begin{equation}
\log \frac{M}{m_b} \to 
\log \frac{M}{m_b + \delta m_b} =
\log \frac{M}{m_b} - \log \left(1 + \frac{\delta m_b}{m_b} \right) \simeq
\log \frac{M}{m_b} - \frac{\delta m_b}{m_b} \; .
\label{eq:mb_log}
\end{equation}
For well motivated applications of the bottom parton densities ($M \gg
m_b$) the uncertainly due to the bottom mass becomes increasingly
irrelevant. For example, a logarithm of the order $\log 100 \sim 4.6$
reduces the relative error from the bottom mass input by this factor
$1/4.6$.

\begin{table}[t]
\begin{small} \begin{center}
\begin{tabular}{c||l|l|l|l||l|l||l|l}
&\multicolumn{4}{c||}{$m_H = 200$~GeV}
&\multicolumn{2}{c||}{$m_H = 500$~GeV}
&\multicolumn{2}{c}{$m_H = 800$~GeV} \\
\hline
&\multicolumn{2}{c|}{7~TeV}
&\multicolumn{2}{c||}{14~TeV}
&\multicolumn{2}{c||}{14~TeV}
&\multicolumn{2}{c}{14~TeV}\\
\hline
$m_b$ &
$\sigma [{\rm pb}]$ & $\sigma/\sigma_{4.75}$ &
$\sigma [{\rm pb}]$ & $\sigma/\sigma_{4.75}$ &
$\sigma [{\rm pb}]$ & $\sigma/\sigma_{4.75}$ &
$\sigma [{\rm pb}]$ & $\sigma/\sigma_{4.75}$ \\
\hline
4.25 
& 0.1845 & 1.055 & 1.279   & 1.049 
& 0.1168 & 1.045 & 0.01989 & 1.044 \\
4.50 
& 0.1796 & 1.026 & 1.248   & 1.025 
& 0.1142 & 1.021 & 0.01945 & 1.021 \\
4.75 
& 0.1750 & 1.0   & 1.219   & 1.0   
& 0.1118 & 1.0   & 0.01905 & 1.0 \\
5.00 
& 0.1708 & 0.976 & 1.192   & 0.978 
& 0.1096 & 0.980 & 0.01868 & 0.981 \\
5.25 
& 0.1668 & 0.953 & 1.166   & 0.957 
& 0.1074 & 0.961 & 0.01832 & 0.977 
\end{tabular}
\end{center} \end{small}
\vspace*{-5mm}
\caption{Absolute and relative production rates for $t H^-$
  production at NLO, varying the input bottom mass in the on-shell
  scheme.  The coupling is fixed by $\tan\beta=30$, and the
  renormalization scale is $\mu = (m_t+m_H)/2$.}
\label{tab:xsmb}
\end{table}

Because in the matrix element squared of the hard process the bottom
mass only enters power suppressed we set it to zero in that part of
the calculation. This includes the collinear divergence which we
regularize in dimensional regularization and move into the parton
densities.  It is disputable if together with the input bottom mass,
{\sl i.e.}  the threshold of the gluon splitting, we should also vary
the bottom Yukawa coupling. However, for large enough $\tan\beta$ the
change of the production rate with the bottom Yukawa coupling is
trivially given by $\sigma \propto y_b^2$. If we evaluate the bottom
Yukawa in the $\overline{\rm MS}$ scheme, changing the on-shell bottom
mass by $\pm 0.5$~GeV changes $y_b(M)$ by roughly $\pm 13\%$ and the
Higgs cross section by $\pm 25\%$.\footnote{Note that in the MC\@@NLO
  implementation the starting value $m_b(m_b)=4.23$~GeV is fixed,
  corresponding to an on-shell value of 4.87~GeV. Varying the input
  bottom mass will not actually change the bottom Yukawa coupling.}

In comparison, varying the bottom mass in the parton densities over
$4.25 \cdots 5.25$~GeV for 7~TeV collider energy leads to a $\pm 5\%$
window for a Higgs mass of 200~GeV, as shown in
Table~\ref{tab:xsmb}. The change in rate is half of the change in the
bottom mass, roughly in line with Eq.(\ref{eq:mb_log}). For larger
collider energies and for larger Higgs masses the relative change in
rate slightly decreases, again as expected. This again confirms
Eq.(\ref{eq:mb_log}) and its conclusion that the effect of the bottom
mass through the parton densities is strongly suppressed in comparison
with the shift in the bottom Yukawa.

\section{Small Higgs masses}

For small Higgs masses a problem arises in the NLO corrections to the
process $bg \to tH^-$: for example the $\mathcal{O}(y_t^2 \alpha_s^2)$
subprocess
\begin{equation}
 gg/q\bar{q} \to t H^- \bar{b} 
\label{eq:sub1}
\end{equation}
includes diagrams with on-shell $gg \to t\bar{t}$ production and a
subsequent decay $\bar{t} \to H^- \bar{b}$. They only include a
two-particle production phase space, $\sigma_{t \bar{t}} \times {\rm
  BR}_{\bar{t} \to H^- \bar{b}}$, and therefore numerically dominate.

From a field theory perspective there exists no problem with these
on-shell diagrams, aside from the usual complications of intermediate
on-shell particles. To compute LHC rates and distributions we can
introduce a width $\Gamma_t$ or apply more advances schemes in
analogy to four fermion production at LEP~\cite{four_f}. From Tevatron
top analyses we expect that the introduction of a width while keeping
the spin correlation of the top quark will be completely sufficient.

The two-fold reason for separating top pairs from top-Higgs production
is phenomenological: first, the two processes have different kinematic
features and therefore require different analyses based on different
(simulated) event samples. Second, the QCD corrections to the two
processes should not be expected to be identical. Therefore, we first
simulate top pair production including QCD effects and normalized to
the measured rate. Top-Higgs production then corresponds to an event
sample which we can add to the top pair sample without any double
counting.

The appropriate treatment of such intermediate states in different $2
\to 2$ production channels has been developed in
PROSPINO~\cite{prospino} and the NLO $pp \to H^- t$
calculation~\cite{mine2}. It was then independently applied to single
top production in MC\@@NLO as so-called diagram
subtraction~\cite{mcnlo_single}. The challenge is to define local
subtraction terms which remove the on-shell process $pp \to t\bar{t}$
from the real emission contribution Eq.(\ref{eq:sub1}). Interference
between the two processes of course counts towards new physics, {\sl
  i.e.} top-Higgs production. This is why diagram removal schemes on
the amplitude level are not useful.

\begin{figure}[t]
\begin{center}
\includegraphics[width=0.46\hsize]{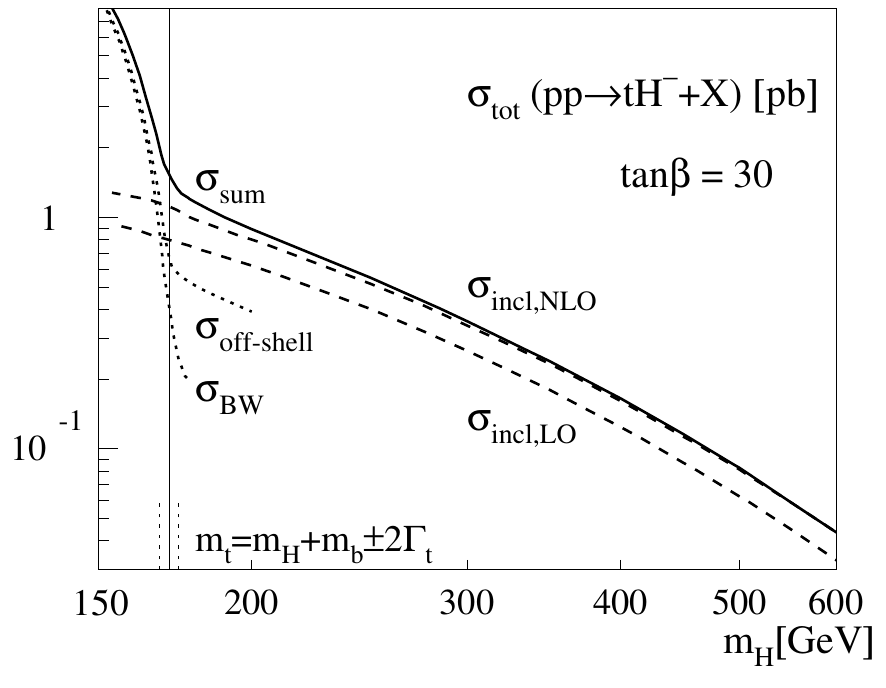}
\hspace*{0.10\hsize}
\includegraphics[width=0.358\hsize]{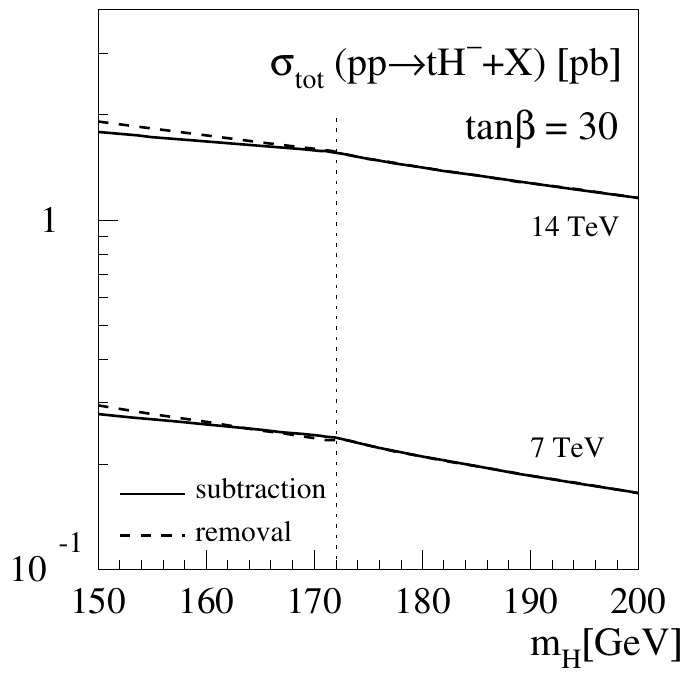}
\end{center}
\vspace*{-8mm}
\caption{Total cross sections for top-Higgs production at the LHC
  including small Higgs masses. Left: PROSPINO
  implementation~\cite{mine2}; Right: MC\@@NLO
  implementation~\cite{mcnlo_th}, for the correct diagram subtraction
  and the inconsistent diagram removal.}
\label{fig:sub}
\end{figure}

In terms of the momentum squared $s_{bh}$ flowing through the
intermediate top propagator and the corresponding residue
$\matx(s_{bh})$ we add a subtraction term to the otherwise unmodified
Breit-Wigner top propagator
\begin{equation}
 \frac{\matx(s_{bh})}{(s_{bh}-m_t^2)^2+m_t^2\Gamma_t^2} 
 \, - \, \frac{\matx(m_t^2)}{(s_{bh}-m_t^2)^2+m_t^2\Gamma_t^2}
 \; \Theta( \hat{s} - 4 m_t^2)
 \; \Theta( m_t - m_H ) \; ,
\label{eq:sub2}
\end{equation}
including the appropriate constrained phase
space~\cite{prospino}. This expression holds independently of the
value of the regulator $\Gamma_t$.  In the PROSPINO implementation
with quasi-stable supersymmetric particles we use the gauge invariant
narrow width limit to uniquely define the subtraction term
\begin{equation}
\lim_{\Gamma_t \ll m_t}
 \frac{m_t \Gamma_t}{(s_{bh}-m_t^2)^2+m_t^2\Gamma_t^2} 
= \pi \; \delta (s_{bh}-m_t^2) \; .
\end{equation}
In contrast, the MC\@@NLO implementation includes top decays, so we
identify the regulator with the physical anomalous top width.

In Figure~\ref{fig:sub} we show the different contributions to
associated top-Higgs production at the LHC~\cite{mine2}. Top pair
production including a Breit-Wigner propagator and the appropriate
branching ratio is labelled $\sigma_{\rm BW}$. It dominates for small
Higgs masses and quickly becomes numerically irrelevant in the
threshold region. This description might be improved by including all
off-shell diagrams contributing to $t H^- \bar{b}$
production. However, $\sigma_{\rm off-shell}$ only shows an effect
above threshold, where it corresponds to a subset of the NLO top-Higgs
diagrams.  Instead, we use the complete subtracted $\sigma_{\rm incl,
  NLO}$ which together with the on-shell top pair production gives the
full rate $\sigma_{\rm sum}$. The resulting $K$ factor for the
subtracted $t H^-$ production is flat across the threshold $m_H =
m_t$.

As a side remark, the four-flavor prediction for $p \to \bar{b} t
H^-$~\cite{charged_bth} includes the on-shell top contribution to
leading order, regularized by the physical top width. However, the
separation into two distinctly different processes would work the same
way, for example if one would want to normalize the top-pair
production rate to data.

\section{Outlook}

\begin{table}[t]
\begin{small} \begin{center}
\begin{tabular}{c||c|c|c||c|c|c|}
&\multicolumn{3}{c||}{$\sigma_{\rm NLO}$ (7~TeV)}
&\multicolumn{3}{c|}{$\sigma_{\rm NLO}$ (14~TeV)}\\
\hline
$m_H$ &
$\mu = \mu_0$ & $\mu = \mu_0/3$ & $\mu = 3 \mu_0$ & 
$\mu = \mu_0$ & $\mu = \mu_0/3$ & $\mu = 3 \mu_0$ \\
\hline
200  & 0.1777  & 0.1944  & 0.1605  & 1.237  & 1.398  & 1.133  \\
250  & 0.1036  & 0.1101  & 0.0934  & 0.7911 & 0.8752 & 0.7232 \\
300  & 0.06266 & 0.06562 & 0.05618 & 0.5212 & 0.5707 & 0.4743 \\
350  & 0.03887 & 0.04027 & 0.03467 & 0.3516 & 0.3813 & 0.3185 \\
400  & 0.02467 & 0.02533 & 0.02189 & 0.2421 & 0.2607 & 0.2184 \\
450  & 0.01598 & 0.01627 & 0.01410 & 0.1698 & 0.1818 & 0.1527 \\
500  & 0.01053 & 0.01065 & 0.00925 & 0.1211 & 0.1290 & 0.1085 
\end{tabular}
\end{center} \end{small}
\vspace*{-5mm}
\caption{NLO Production rates in pb for $pp \to t H^-$ production in
  the five-flavor scheme, using the input parameters defined by the LHC
  Higgs cross sections working group. The central renormalization and
  factorization scales are chosen as $\mu_0 = (m_t +
  m_H)/4$~\cite{mine1,mine2}. The bottom Yukawa scales with $\tan
  \beta = 30$.}
\label{tab:xsref}
\end{table}

We have reviewed the status of theoretical predictions for the
production rate of associated $t H^-$ production in a two-Higgs
doublet model. Two schemes describe bottom initiated processes at
hadron colliders, with or without bottoms densities inside the
proton. To all orders in perturbative QCD they are equivalent, but to
leading order they give different results. Both schemes have been
implemented to NLO, and we find reasonable agreement for the
production rates.

For distributions the difference between these two schemes is not the
main issue. The relevant collinear bottom and light-flavor jet
radiation is be poorly described by either of them and requires a
combination with a parton shower. MC\@@NLO describes the kinematics of
bottom and light-flavor jet radiation reliably over the entire phase
space and allows us to study the features of radiated bottom jets,
radiated light-flavor jets, and jets from the top decay.

One yet unstudied issue in the five-flavor scheme is the dependence of
the rate on the bottom mass assumed for the bottom parton
densities. The effect of a shift in the input bottom mass is
suppressed by the same collinear logarithm which justifies the use of
bottom parton densities. Its numerical impact does not exceed a few
per-cent.

For light charged Higgs bosons it is crucial that we consistently
combine higher order corrections of the kind $pp \to t H^- \bar{b}$
with top pair production followed by an anomalous decay $\bar{t} \to
H^- \bar{b}$.  There exists a unique prescription how to separate the two
processes using a point-wise phase space subtraction. We have reviewed
the basic features of this PROSPINO subtraction scheme and its
MC\@@NLO implementation.

Finally, we give the NLO cross section predictions
for the five-flavor scheme using the input parameters defined by the
LHC Higgs cross section working group. For two different LHC collider
energies and a range of renormalization and factorization scales they
are shown in Table~\ref{tab:xsref}.\bigskip

We would like to thank the organizers of this workshop series for the
very inspiring and constructive atmosphere. Moreover, we would like to
thank Michael Kr\"amer for many constructive discussions.

\end{document}